\documentclass{PoS}

\usepackage{amsfonts}
\usepackage{amssymb}
\usepackage{amsmath}
\usepackage{cancel}

\usepackage{graphicx}
\usepackage{dcolumn}
\usepackage{bm}
\usepackage{epsfig}
\usepackage{psfrag}
\usepackage{sidecap}

\usepackage{dsfont}
\usepackage{cancel}
\usepackage{graphicx}

\title{Electroweak hadron structure in point-form dynamics -- heavy-light systems}

\ShortTitle{Electroweak hadron structure in point form dynamics}

\author{\speaker{Mar\'ia G\'omez-Rocha}\thanks{Supported by Fonds zur F\"orderung der wissenschaftlichen Forschung in \"Osterreich (FWF DK W1203-N16)}~ and Wolfgang Schweiger\\
       Institut f\"ur Physik, Universit\"at Graz, A-8010 Graz, Austria\\
        E-mail: \email{maria.gomez-rocha@uni-graz.at}, \email{wolfgang.schweiger@uni-graz.at}}

\abstract{We present a general formalism that uses the point form of relativistic Hamiltonian dynamics
to describe the electroweak structure of heavy-light mesons within constituent quark models.
We study the heavy quark limit (i.e. $m_Q\to \infty$) and check that the predictions of heavy quark symmetry are satisfied.
A simple analytic expressions is given for the Isgur-Wise function. In addition, cluster properties and the relation of our approach to front form calculations are discussed.}

\FullConference{Sixth International Conference on Quarks and Nuclear Physics,\\
		April 16-20, 2012\\
		Ecole Polytechnique, Palaiseau, Paris}

\begin{document}

\section{Introduction}
In his seminal paper of 1949 \cite{Dirac:1949cp} Dirac proposed the point form as one of three prominent ways to formulate relativistic Hamiltonian dynamics. It is the least explored form of relativistic dynamics, although its has several properties that make it very convenient in intermediate energy hadron problems. Unlike the instant and the front form, the point form has, e.g., the nice feature that boost and rotation generators are not affected by interactions, they are kinematical.
This allows to transform bound states in a simple way. In turn, all 4 components of the 4-momentum operator become interaction dependent. Our form-factor calculations are based on the point form of relativistic quantum mechanics \cite{Keister:1991sb}. We use the Bakamjian-Thomas construction for introducing interactions in our system~\cite{Keister:1991sb,Bakamjian:1953kh} such that Poincar\'e invariance is guaranteed. This approach has already successfully been applied to the study of electromagnetic properties of spin-0 and spin-1 two-body bound states consisting of equal-mass particles \cite{Biernat:2009my,Biernat:2010tp,Biernat:2011mp}. The equivalence with results obtained in a covariant light-front formulation~\cite{Carbonell:1998rj} has been established~\cite{Biernat:2011mp,GomezRocha:2011qs}. Most recently this formalism was extended to the study of the electroweak structure of mesons consisting of constituents with different masses~\cite{Rocha:2010wm,GomezRocha:2012zd}. The heavy-quark symmetry properties of heavy-light mesons were tested for the case that the mass of the heavy quark goes to infinity. We are now able to compute electromagnetic and weak (decay) form factors of pseudoscalar and vector mesons. In the present work we want to consider in some more detail the heavy-quark limit of the electromagnetic current of a pseudoscalar heavy-light meson. We will show, in particular, how problems with cluster separability can be handled and how our results are related to front form calculations.

\section{Transition amplitudes, currents and form factors of electroweak processes}
We start from the physical processes in which electroweak form factors of heavy light systems can be measured:
electromagnetic scattering and weak decays of heavy-light mesons. These processes are treated within a coupled channel approach to account for the dynamics of the exchanged $\gamma$ or $W$. For a
Poincar\'e invariant formulation of these reactions we use the Bakamjian-Thomas construction~\cite{Bakamjian:1953kh}. Its
point-form version takes on the form:
\begin{equation}
 \hat P^\mu=\hat M \hat V^\mu_{\text{free}}=(\hat M_{\text{free}}+\hat M_{\text{int}})
\hat V^\mu_{\text{free}}\, .
\end{equation}
This means that the (interacting) 4-momentum operator is factorized into an interacting mass operator and
a free 4-velocity operator. One thus has to study only 1 eigenvalue problem for the mass operator $\hat{M}$ and not 4 simultaneous eigenvalue problems for the components of the 4-momentum operator $\hat{P}^\mu$. Electron-meson scattering, e.g., is now formulated on a Hilbert space that consists of a $q\bar{q}e$ and a $q\bar{q}e\gamma$ sector. A convenient basis for this Hilbert space is given by
\textit{velocity states} which characterize the state of a multiparticle system by its overall velocity $V$ and the
center-of-mass momenta and spins of its components~\cite{Klink:1998zz}.
The mass eigenvalue equation that has to be solved is
\begin{eqnarray}\label{mass:eq}
 \left(\begin{array}{cc} \hat M_{eq\bar q}^{\text{conf}} & \hat K \\ \hat K^\dagger & \hat M_{eq\bar q\gamma}^{\text{conf}}\end{array}\right)
 \left(\begin{array}{c}  |\psi_{eq\bar q}\rangle \\   |\psi_{eq\bar q\gamma}\rangle \end{array}\right)   =
 m \left(\begin{array}{c} |\psi_{eq\bar q}\rangle \\ |\psi_{eq\bar q\gamma}\rangle \end{array}\right)\, .
\end{eqnarray}
The diagonal elements of the matrix mass operator contain the relativistic kinetic energies and
an instantaneous confining interaction between the quark and the antiquark. $\hat K^\dagger$ and $\hat K$ are
vertex operators which describe the emission and absorption of the photon by quark, antiquark, or electron. They
are related to the usual interaction Lagrangian density of QED~\cite{GomezRocha:2012zd}.
Eliminating the $q\bar{q}e\gamma$ channel, which only plays a role in the intermediate state, we end up with an equation for the $q\bar{q}e$ component of the mass eigenstate
\begin{equation}
 (\hat M^{\text{conf}}_{eq\bar q}-m)|\psi_{eq\bar q}\rangle =
\underbrace{\hat K (\hat M^{\text{conf}}_{eq\bar q\gamma}-m)^{-1}\hat K^\dagger}_{\hat V_{opt}(m)}|\psi_{eq\bar q}\rangle\, .
\end{equation}
The optical potential $\hat V_{opt}(m)$ describes the 1-photon exchange between electron and (anti)quark.
On-shell matrix elements of the optical potential between (velocity) states of a confined $q\bar{q}$ pair with quantum numbers of the meson $M$ provide the invariant 1-photon-exchange amplitude from which the electromagnetic current of the meson $M$ can be extracted. As one would expect, these matrix elements can be written as a contraction of the electromagnetic electron current with an electromagnetic hadron current (spin projections $\mu_M^{(\prime)}$ and $\mu_e^{(\prime)}$ are suppressed):
\begin{equation}
 \langle V'; \vec k_e';\vec k_M',\mu^\prime_M|\hat V_{\text{opt}}(m)|V;\vec k_e;\vec k_M,\mu_M\rangle_{\text{on-shell}}
\propto V^0\delta^3(\vec V -\vec V')\frac{j_\mu(\vec k_e';\vec k_e)J^\mu(\vec k_M';\vec k_M)}{(k_e'-k_e)^2}.
\end{equation}
This fixes the hadron current and hence the electromagnetic hadron form factors in a unique way. Transition amplitudes for semileptonic weak decays and weak transition currents can be calculated analogously~\cite{GomezRocha:2012zd}.

The electromagnetic current obtained in this way transforms like a 4-vector (if it is reexpressed in terms of physical hadron momenta) and it is conserved in the case of pseudoscalar mesons. It, however, exhibits unwanted cluster properties.
This is the price we have to pay for the Poincar\'e invariance of our formulation. It is a well known problem of the Bakamjian-Thomas construction~\cite{Keister:1991sb}. In our case the electromagnetic current cannot be covariantly decomposed in terms of the incoming and outgoing hadron 4-momenta alone, one needs also the electron 4-momenta.\footnote{This resembles the situation in the covariant front-form approach of Carbonell et al.~\cite{Carbonell:1998rj} in which the currents depend on a 4-vector that specifies the orientation of the light front.} For the electromagnetic
current of a pseudoscalar meson one gets (see \cite{Biernat:2011mp,GomezRocha:2011qs} for vector mesons):
\begin{equation}\label{J:f:g}
 J^\mu(\vec k'_M;\vec k_M)= (k_M+k_M')^\mu f(Q^2, k)+ (k_e+k_e')^\mu g(Q^2, k)\, .
\end{equation}
Violation of  cluster separability affects also the form factors which exhibit, in addition to the usual $Q^2=-q_\mu q^\mu$ dependence, a dependence on  the modulus of the meson CM-momentum $k:=|\vec k_M|$ (or equivalently Mandelstam-$s$).
Numerical and analytical studies show that the spurious dependencies of the electromagnetic current on the electron momenta  vanish rather fast with increasing $k$ for pseudoscalar mesons. Remarkably, the $k\to\infty$ limit of the electromagnetic form factor then turns out to agree with the front-form result computed in the $q^+=0$ frame~\cite{Biernat:2009my,Biernat:2010tp,Biernat:2011mp}.

In the following section we will elucidate whether and to which extend the effects of wrong cluster properties still play a role for the electromagnetic current~(\ref{J:f:g}) of a heavy-light pseudoscalar meson  when the heavy-quark limit $m_Q\rightarrow \infty$ is performed.

\section{The heavy-quark limit}
\begin{figure}[t!]
\includegraphics[width=0.5\textwidth]{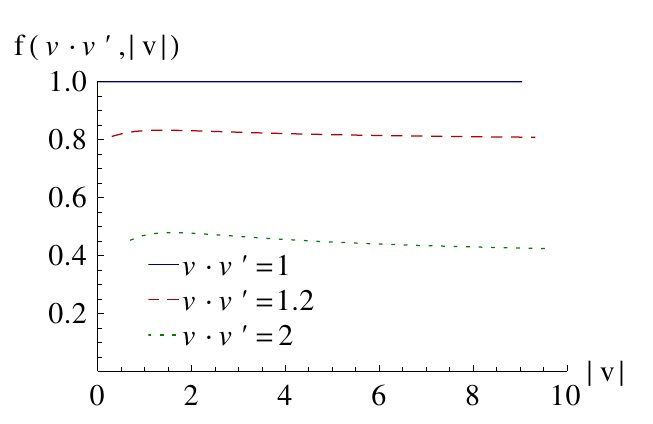}
\includegraphics[width=0.5\textwidth]{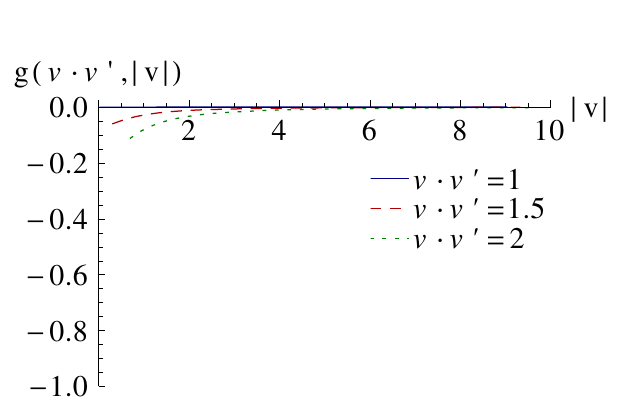}
\caption{$|\vec v|$-dependence of the physical and spurious
$B^-$ electromagnetic form factors $\tilde f(v\cdot v',|\vec v|)$ and
$\tilde g(v\cdot v',|\vec v|)$, respectively, for different values of $v\cdot v'$.
Shown is the heavy-quark limit. Calculations are done with a simple harmonic-oscillator wave
function (see Ref.~\cite{GomezRocha:2012zd}) with oscillator parameteter $a=0.55$~GeV. The light-quark mass is $m_{\bar u}=0.25$ GeV.}\label{fig:vdep}
\end{figure}

In the heavy-quark limit (h.q.l.) the masses of the heavy quarks (and therefore the heavy-meson masses)
go to infinity ($m_M=m_Q \to \infty$). As a consequence, the meson momenta go to infinity as well and it is thus more appropriate to use the meson velocities.
The 4-momentum transfer squared, which goes to infinity too, is then replaced by
the product of the initial and final 4-velocities $v\cdot v'$ of the meson. For elastic electron-meson scattering one has $q^2=2m_M^2(1-v\cdot v')$. The heavy-quark limit has to be taken in such a way
that the product $v\cdot v'=k_M\cdot k_M'/m_M^2$ stays constant.
After separation of $m_M$ the electromagnetic current~(\ref{J:f:g}) (expressed in terms of velocities) becomes in the h.q.l.
\begin{equation}\label{eq:jemhqcov}
\frac{1}{m_M}\, J^\mu(\vec k'_M,\vec k_M)\quad \stackrel{\mathrm{h.q.l.}}{\longrightarrow}\quad {J}^{\mu}_\infty
(\vec{v}^\prime,\vec{v}) =
(v+v^{\,\prime})^\mu
\,\tilde{f}(v\cdot
v^{\prime},|\vec v|) +
\frac{m_e}{m}(v_e+v_e^{\,\prime})^\mu
\,\tilde{g}(v\cdot
v^{\prime},|\vec v|)\, .
\end{equation}
Since calculations are done in the electron-meson CM system $v_e^{(\prime)}=\mathcal{O}(m_M)$ and thus the second covariant is not negligible as compared to the first one. The $Q^2$ and $|\vec{k}|$ dependencies of the form factors go over in $v\cdot
v^{\prime}$ and $|\vec{v}|$ dependencies, respectively. Numerical studies reveal that the spurious contributions caused by cluster problems become small in the heavy-quark limit, but they do not vanish (see Fig.~\ref{fig:vdep}). There are, however, two interesting limiting cases in which these contributions do not play a role.

\subsection{The infinite-momentum frame}
With \lq\lq infinite-momentum frame\rq\rq~(IF) we mean the situation in which $|\vec v|\to \infty$ (after the heavy-quark limit has been performed). In the $|\vec v|\to \infty$ limit the unwanted $|\vec v|$-dependence goes away and the spurious form factor vanishes such that
\begin{equation}\label{covariant:if}
 \tilde J^\mu_\infty (\vec v' ,\vec v)\stackrel{|\vec v|\rightarrow\infty}{\longrightarrow}
(v+v')^\mu \xi_{IF} (v \cdot v')\, .
\end{equation}
We identify $\xi_{IF} (v \cdot v')$ as the Isgur-Wise function. Its analytical expression is
rather simple
\begin{equation}\label{IW:IF}
 \xi_{IF}(v\cdot v')=\int \frac{d^3 \tilde k'_{\bar q}}{4\pi}
\sqrt{\frac{\tilde k^0_{\bar q}}{\tilde k'^0_{\bar q}}}
\mathcal S_{IF}\psi^*(|\vec{\tilde k}'_{\bar q}|)\psi(|\vec{\tilde k}_{\bar q}|), \;\;
\text{with spin factor} \;\;\mathcal S_{IF}=\frac{m_{\bar q}+\tilde k'^0_{\bar q}+\tilde k'^1_{\bar q}u}
{\sqrt{(m_{\bar q}+\tilde k^0_{\bar q})(m_{\bar q}+\tilde k'^0_{\bar q})}},
\end{equation}
$2u^2=v\cdot v'-1$. $\tilde{k}_{\bar{q}}$ and $\tilde{k}_{\bar{q}}^\prime$ are related by a Lorentz boost~\cite{GomezRocha:2012zd}.

\subsection{The Breit frame}
\begin{SCfigure}
\includegraphics[width=0.5\textwidth]{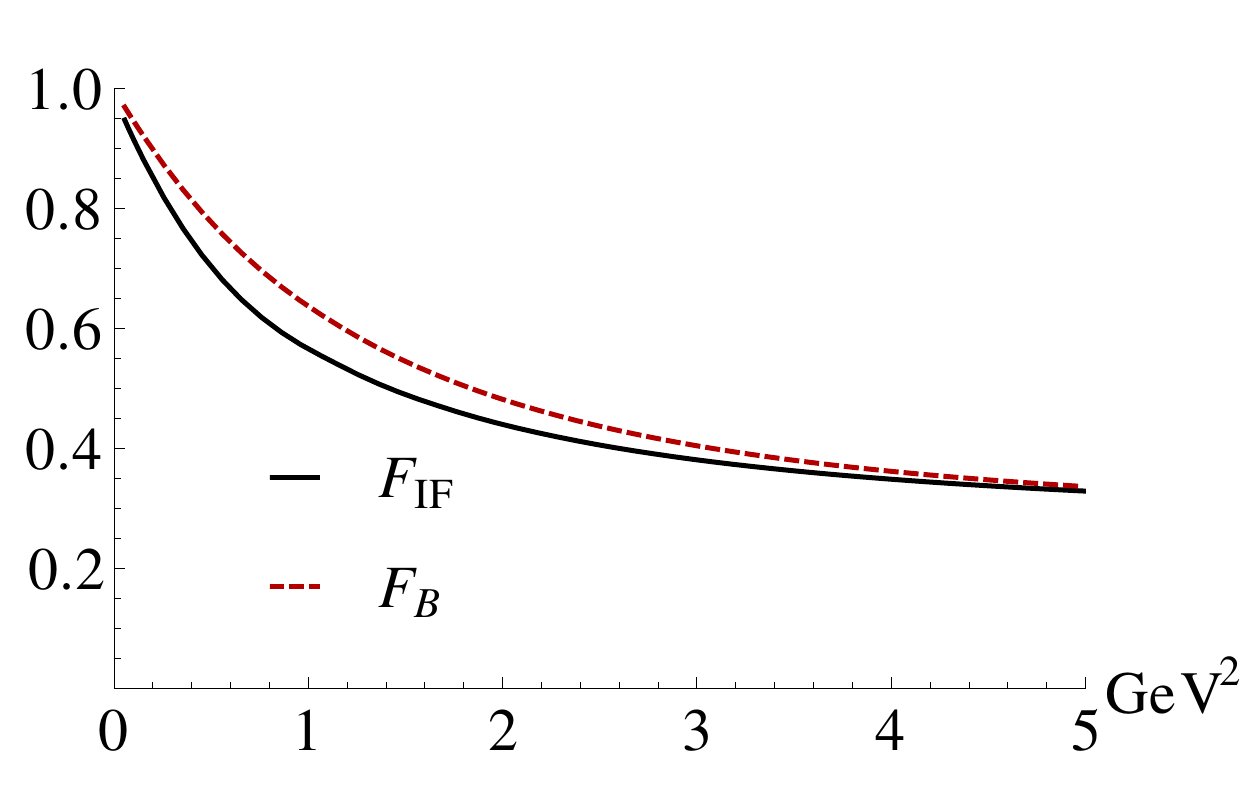}
\caption{
Electromagnetic form factor of the $B^-$
calculated in the Breit (B) and infinite-momentum (IF) frames.
Model parameters are the same as in Fig.~1 and
$m_b=4.8$ GeV, $m_B=5.279$ GeV.
}\label{BQ2BreitInf}
\end{SCfigure}

In the \lq\lq Breit frame\rq\rq\ the energy transfer between the meson in the initial and the final state vanishes.
This is just the opposite situation to the infinite-momentum frame, since $|\vec v|$ takes now
the minimal value ($|\vec v|^2=u^2=(v \cdot v'-1)/2$). In this case the physical and spurious covariant become proportional and the corresponding form factors cannot be separated
\begin{equation}
 \tilde J_\infty^\mu(\vec v',\vec v')\stackrel{|\vec v|\rightarrow u}{\longrightarrow}
(v+v')^\mu \left\{ \tilde f(v \cdot v',|\vec v|=u)+\sqrt{\frac{v\cdot v'-1}{v\cdot v'+1}}\tilde g(v \cdot v',|\vec v|=u)\right\} =:(v+v')^\mu \xi_B(v\cdot v').
\end{equation}
The covariant structure is the same as in Eq.~(\ref{covariant:if}) and
also the analytical form of the invariant function $\xi_B (v\cdot v')$ looks quite similar:
\begin{equation}\label{IW:breit}
 \xi_{B}(v\cdot v')=\int \frac{d^3 \tilde k'_{\bar q}}{4\pi}
\sqrt{\frac{\tilde k^0_{\bar q}}{\tilde k'^0_{\bar q}}}
\mathcal S_{B}\psi^*(|\vec{\tilde k}'_{\bar q}|)\psi(|\vec{\tilde k}_{\bar q}|), \;\;
\text{with spin factor} \;\; \mathcal S_{B}=\frac{m_{\bar q}+\tilde k'^0_{\bar q}+\tilde k'^1_{\bar q}\frac{u}{\sqrt{u^2+1}}}
{\sqrt{(m_{\bar q}+\tilde k^0_{\bar q})(m_{\bar q}+\tilde k'^0_{\bar q})}}.
\end{equation}
It can indeed be proved analytically that Eqs.~(\ref{IW:IF}) and (\ref{IW:breit}) are equivalent and related by a simple rotation of the integration variables~\cite{GomezRocha:2012zd}. The result for the Isgur-Wise function thus turns out to be independent on whether it is extracted in the Breit or in the infinite-momentum frame. This is a particular feature of the
heavy-quark limit and does not hold for finite meson masses. As an example Fig.~\ref{BQ2BreitInf} shows the electromagnetic form factor of the $B^-$ meson, calculated in the infinite-momentum and Breit-frame, respectively.

Weak decays of heavy-light mesons can be treated in an analogous way. In weak decays, however, one does not observe any manifestation of wrong cluster properties via spurious dependencies of the current~\cite{GomezRocha:2012zd}. Actually, the kinematics of the weak decay process is close to the Breit frame kinematics in electron-meson scattering -- with the only difference that we are now dealing with a time-like process. The Isgur-Wise function one obtains in the heavy-quark limit from the decay form factors is indeed identical to (\ref{IW:breit}), proving that the features of heavy-quark symmetry~\cite{Isgur:1989vq,Isgur:1989ed,Neubert:1993mb} emerge in our approach.

\section{Comparison with front form calculations}
Like the electromagnetic form factors of pseudoscalar mesons (extracted in the infinite-momentum frame) also the Isgur-Wise function resulting from our approach is found to agree numerically with corresponding front-form calculations~\cite{Cheng:1996if}. For finite quark masses, however, differences show up in the weak $B\rightarrow D^{(\ast)}$ decay form factors. Some frame dependence, which we don't get, is in particular observed for the $B\to D^*$ decay form factors when they are extracted from the +-component of a one-body current~\cite{Cheng:1996if,Bakker:2003up}. This is attributed to a missing $Z$-graph (non-valence) contribution which is suppressed in a $q^+=0$ frame, but should be included in other frames to preserve the covariance properties of the current. This affects, in particular the decay current, since the $q^+=0$ condition cannot be imposed for decays. In the case of the point form it is also not excluded that Z-graphs may play a role, but they are not necessary to ensure covariance of the current, since Lorentz boosts are purely kinematical and thus do not mix in higher Fock states. Nevertheless, it may be interesting to see, whether the inclusion of Z-graph contributions in our framework could help to reduce the spurious dependencies of the electromagnetic current on the electron momenta. $Z$-graphs, could be easily accommodated within our multichannel approach and it will be the topic of future work to see, whether they could, e.g., explain the discrepancy between the $B^-$ electromagnetic form factors calculated in the infinite-momentum and Breit frame, respectively (see Fig.~2).

\end{document}